\documentclass[peerreview]{IEEEtran}
\IEEEoverridecommandlockouts
\usepackage{cite}
\usepackage{amsmath,amssymb,amsfonts}
\usepackage{algorithmic}
\usepackage{graphicx}
\usepackage{textcomp}
\usepackage{xcolor}
\usepackage[hidelinks]{hyperref}
\usepackage{multirow}
\usepackage{comment}
\usepackage{mathptmx}

\newcommand{\purple}[1]{\textcolor{black}{#1}}

\begin{document}

\title{Reading the Eyes in VR: Multimodal Modeling of Social Intelligence}

\author{
\IEEEauthorblockN{Mohammad Fahim Abrar}
\IEEEauthorblockA{\textit{Computer and Information Science} \\
\textit{University of Delaware}\\
Newark, Delaware, USA \\
fahim@udel.edu}
\and
\IEEEauthorblockN{Shyala Sharmin}
\IEEEauthorblockA{\textit{Computer and Information Science} \\
\textit{University of Delaware}\\
Newark, Delaware, USA \\
shayla@udel.edu}
\and
\IEEEauthorblockN{Roghayeh Leila Barmaki}
\IEEEauthorblockA{\textit{Computer and Information Science} \\
\textit{University of Delaware}\\
Newark, Delaware, USA \\
rlb@udel.edu}
}

\maketitle

\maketitle

\begin{abstract}
Virtual reality (VR) provides an immersive setting for social-cognitive assessment and multimodal sensing, but the predictive value of behavioral and physiological signals and the experience of delivering the same assessment in VR versus conventional desktop settings remain important empirical questions. In this proof-of-concept study, we investigate whether combining synchronized electroencephalography (EEG) and gaze improves observed predictive information for modeling performance on the Reading the Mind in the Eyes Test (RMET), alongside an exploratory comparison of VR and desktop delivery. We recruited 20 participants, with 16 providing usable data in a between-subjects comparison of VR ($n=8$) and desktop ($n=8$) conditions. We evaluated question-level correctness classification and participant-level score regression, including EEG-only, gaze-only, and combined representations for regression. In participant-level regression, combined EEG--gaze representations yielded lower prediction errors than either modality alone. The best observed model achieved $R^2=0.63$ and a mean absolute error of 1.68 RMET points. The VR group reported higher usability and effort, while mean fixation time per question did not differ significantly between groups. These preliminary findings provide proof of concept for multimodal AI-based modeling of task-specific RMET performance across VR and desktop recordings. 
\end{abstract}


\section{Introduction}

Virtual reality (VR) provides controlled, immersive environments for cognitive assessment and social-cognitive training~\cite{kourtesis2021validation,vass2022virtual}. These applications offer opportunities to study how people interpret information and respond to task demands. A central challenge is how to connect conventional assessment outcomes with the behavioral and physiological signals generated during task completion. Correctness and total scores summarize task success, but do not directly describe participants' visual exploration or physiological responses. This motivates investigating whether signals recorded during assessment contain information that can support AI-based modeling of task performance.

Multimodal sensing allows researchers to examine how task performance relates to behavioral and physiological signals recorded during task completion. Eye tracking records how users inspect visual stimuli, while electroencephalography (EEG) captures brain activity during the task. Synchronizing these modalities with task events allows machine learning models to examine their predictive value individually and in combination. Prior work has demonstrated benefits from combining EEG and eye-tracking data for emotion recognition~\cite{Zheng2014}, while research on attention classification has shown that the value of multimodal information depends on its representation and fusion~\cite{vortmann2022multimodal}. However, evidence that multimodal fusion benefits one prediction target does not establish that the same representation will be informative for task-specific assessment outcomes. In particular, it remains unclear whether synchronized EEG and gaze provide complementary predictive information for performance on a fixed social-cognitive task when observations are evaluated on held-out participants.

Evaluating multimodal modeling across immersive and conventional settings also requires attention to the setting in which those signals are collected. VR can alter how participants interact with an otherwise comparable task, making the assessment experience relevant to interpreting multimodal data collected across platforms. In a science-learning simulation, immersive VR increased presence while also increasing EEG-derived cognitive load and reducing learning relative to desktop presentation~\cite{makransky2019adding}. The potential benefits of immersive assessment should therefore be evaluated alongside the demands placed on participants. Comparing the same assessment content in VR and on a desktop can characterize differences in gaze behavior, usability, and workload.

The \emph{Reading the Mind in the Eyes Test} (RMET) provides a controlled test case for this investigation. Participants select one of four words to describe the mental or emotional state conveyed by each of 36 photographs of the eye region~\cite{baron2001reading}. Its fixed images and response options allow the same assessment content to be presented through desktop and VR interfaces, with observable outcomes at both the question and total-score levels. We treat RMET performance as a task-specific prediction target, consistent with concerns about interpreting the test as a general measure of theory of mind~\cite{oakley2016theory}. Predicting these outcomes from synchronized recordings allows us to examine whether EEG and gaze contain performance-related information that may be useful for multimodal user modeling across immersive and conventional assessment settings.

Using the RMET as a common assessment task, we investigate the predictive value of EEG and gaze features and the experience of completing the task through VR and desktop interfaces. The study addresses two research questions:

\begin{description}
    \item[\textbf{RQ1:}] To what extent can EEG and gaze features predict RMET performance, measured as (a) question-level correctness and (b) participant-level total score, and for total-score prediction, how do EEG-only, gaze-only, and combined EEG--gaze representations compare?
    \item[\textbf{RQ2:}] How does VR-based delivery of the RMET compare with desktop-based delivery in gaze behavior, perceived usability, and perceived workload?
\end{description}
We conducted an exploratory study with 16 participants using desktop and VR implementations of the same RMET task with synchronized EEG, gaze, and interaction logging. The two implementations preserved the same assessment content while differing in presentation and interaction context; the VR version displayed the task on a virtual screen within a simulated room. This study makes three contributions: (1) a synchronized multimodal pipeline for modeling question-level and participant-level RMET performance from EEG and gaze; (2) a participant-wise evaluation of EEG-only, gaze-only, and combined EEG--gaze representations for total-score prediction using pooled recordings from both settings; and (3) an exploratory between-group characterization of gaze behavior, perceived usability, and workload during VR and desktop delivery. Together, these analyses separate the predictive value of the recorded signals from the experience of the platform through which those signals were collected.

\section{Related Work}
\label{related-work}

\subsection{Assessing Social Cognition: RMET}
The Reading the Mind in the Eyes Test (RMET) is one of the most widely used measures of social-cognitive ability and mental-state inference, requiring participants to infer emotional or mental states from photographs of the eye region \cite{oakley2016theory, higgins2023reading}. It is situated within a broader landscape of theory-of-mind and social-cognitive assessments, including the Strange Stories test, the Faux Pas Recognition Test, MASC, and TASIT \cite{jolliffe1999strange, baron1999new, dziobek2006introducing, mcdonald2006reliability, koo2025}. Its sensitivity is well documented, with individuals with ASD often scoring lower, supporting its relevance for studying differences in social-cognitive processing \cite{oakley2016theory}. The RMET remains especially attractive because it is simple, quiz-based, quick to administer, and has been used across diverse populations \cite{stafford202320, li2022exploring, vellante2013reading}. It focuses on interpreting mental and emotional states from minimal visual cues and has been used in more than 800 studies of social cognition \cite{higgins2023reading}.

At the same time, recent reviews have raised concerns about its psychometric properties, including uncertainty about the specific aspect of social cognition it measures, inconsistent factor structures, relatively low internal reliability, and variation in scores due to demographic factors such as culture, vocabulary, education, and social norms \cite{higgins2024construct}. These concerns have motivated calls for stronger construct-validity evidence, although other work argues that such critiques should be interpreted in context, noting that the RMET continues to show useful group-level sensitivity across both clinical and non-clinical populations and retains practical value when used within its intended scope \cite{murphy2024strong}. From this perspective, we use the RMET in this study as a task for assessing mental-state inference from facial cues and complement behavioral performance with gaze and EEG features that reflect attentional and neural processing.

\subsection{VR for Social Cognition and Multimodal Sensing}

VR is increasingly being used to support social-cognition training across different populations. VR-based interventions for social cognition have shown promising, though mixed, outcomes in schizophrenia and have demonstrated good acceptability and targeted skill improvements in autism, suggesting potential advantages over conventional 2D approaches across clinical populations \cite{vass2022virtual, kourtesis2023virtual}. VR research has also examined nonverbal communication affordances, affective interactions in therapeutic settings, and avatar-based emotional expression, all of which help inform the design of socially expressive and therapeutically relevant virtual environments \cite{deighan2023social, chen2025understanding, ahmadpour2025affective}. Studies have also explored low-latency speech-driven gesture generation for continuous social-XR interaction and the effects of avatar realism on presence and user preference \cite{krome2024minimal, gasch2024avatar}. Work on multimodal communication in VR further suggests that combining channels can improve collaboration \cite{ghamandi2024unlocking}.

At the same time, VR serves as a powerful multimodal sensing platform for studying social cognition under controlled yet realistic conditions. Unlike paper- or screen-based tasks, VR can simulate interactive scenarios that evoke more natural behaviors, including gaze shifts, embodied responses, and emotional reactions \cite{gonzalez2025cognitive}. This strength is closely tied to immersion and presence, which can make virtual experiences feel behaviorally meaningful while preserving experimental control \cite{slater2016enhancing}. VR also supports the integration of eye tracking, EEG, motion capture, and other physiological sensing within the same environment, enabling researchers to capture synchronized behavioral and neural responses during task performance \cite{gonzalez2025cognitive, nolte2025combining}. Prior work has shown that eye tracking in VR is feasible for objective symptom assessment, monitoring engagement, and personalizing therapeutic experiences, which further highlights its value for both assessment and intervention \cite{adhanom2023eye}. Recent studies have likewise explored pupil-dilation signals for predicting cognitive failures in VR and shown that gaze and blink measures can predict task engagement in small-group Social VR interactions \cite{niknam2024predicting, gonzalezdiaz2026gaze}. As a result, VR-based social-cognitive assessment may better approximate real social engagement than traditional static methods \cite{kourtesis2024comprehensive}. However, not all VR assessments exploit dynamic social interaction. Immersive delivery of conventional tasks can instead serve as an intermediate setting for examining multimodal measurement while preserving standardized task content.

Research combining VR with multimodal sensing is already producing promising results for cognitive and affective assessment. Multimodal VR studies that combine gaze, EEG, and behavior have demonstrated the feasibility of predicting affect-related states, supporting the possibility of more automated and continuous assessment \cite{wiebe2024virtual, wiebe2023multimodal}. Studies combining eye tracking with EEG in VR have also supported more naturalistic evaluation pipelines by linking gaze behavior with neural responses \cite{nolte2025combining}. Beyond social cognition, VR tasks integrating multimodal sensing have also shown utility in predicting real-world functioning more effectively than some standard neuropsychological tools \cite{kourtesis2020guidelines}. Together, these findings motivate the use of VR as a platform for multimodal measurement and automated modeling.

\subsection{Machine Learning for Multimodal Cognitive Assessment}

Machine learning with multimodal data has shown strong potential for classifying cognitive and emotional states by combining complementary signals such as EEG, gaze, motion, and behavioral data \cite{lima2024multimodal, tan2017multimodal}. Reviews of immersive computing datasets further highlight the growing foundation for dataset-driven ML in VR/AR/MR research \cite{asish2025synthesizing}. Prior studies have shown that combining EEG and eye tracking can outperform single-modality models in tasks such as emotion recognition and mental workload estimation \cite{lim2020emotion, Zheng2014, Aksu2024}, while more recent VR-based work has extended this approach to multimodal assessment in interactive settings \cite{wiebe2024virtual}. Machine-learning approaches have also been applied to EEG-based VR cybersickness classification using open-access datasets \cite{li2025vr}. The modeling approaches used in this area range from traditional classifiers such as SVMs, random forests, and basic neural networks to more recent deep learning methods including CNNs, LSTMs, and Transformer-based architectures, with the choice of model generally shaped by the scale and complexity of the available data \cite{lim2020emotion, jin2024residual, hazmoune2024using, sundaresan2021evaluating}. Together, these studies suggest that multimodal ML can capture meaningful patterns from heterogeneous signals and motivate its use for cognitive and social-cognitive assessment.
\section{Materials and Methods}
\label{materials-methods}
\subsection{Study Design and Participants}


\subsubsection{Participants}

An a priori power analysis (G*Power) for our between-subjects design (VR vs.\ Desktop) indicated a minimum sample size of $N=16$ was required to achieve a statistical power of $.80$ (assuming a large effect size and $\alpha=.05$). While we initially recruited 20 participants from the university community (10 per condition), recording errors interrupted data collection for four individuals (two per condition), leading to their exclusion. Consequently, our final analyzed sample consisted of 16 participants (eight VR, eight desktop).

Participation was voluntary, and no monetary compensation was provided. Informed consent was obtained from all participants before the study. The study was approved by the University of X’s Institutional Review Board. The final sample consisted of 12 male and four female participants ranging in age from 20 to 33 years ($M=24.93$, $SD=4.10$). Nine participants were graduate students, six were undergraduate students, and one was a postdoctoral researcher. All participants had a background in computer science. Participants reported a mean prior VR experience rating of 3.6 ($SD=1.0$) on a 1--5 scale, where 1 represented ``Not at all familiar'' and 5 represented ``Very familiar.''

\begin{figure}
\centering
  \includegraphics[width=\linewidth]{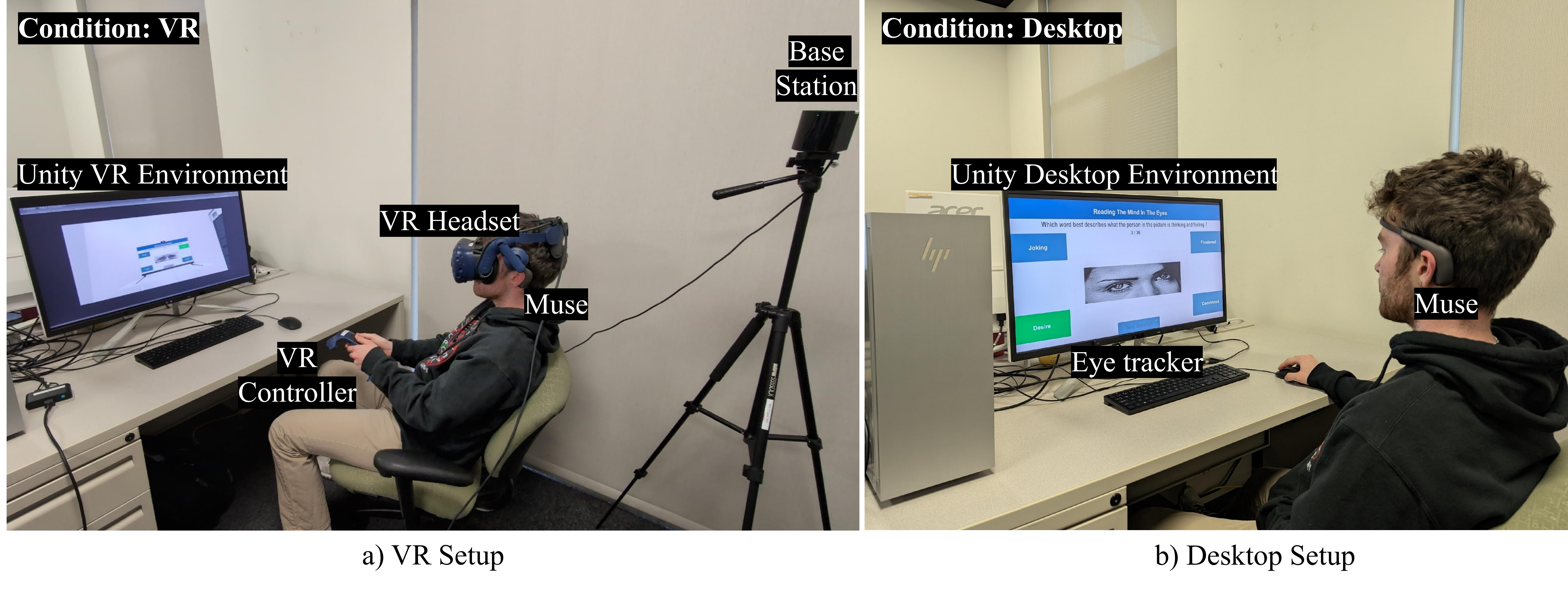}
  \caption{ Experimental setup for the study. (a) VR setup with headset, controllers, base stations, and Muse EEG device. (b) Desktop setup with an eye tracker and Muse EEG device integrated with Unity environments. }
  \label{fig:Study_Sessions}
\end{figure}

\subsubsection{Study Design}

The study used a two-condition between-subjects design in which each participant completed the RMET once, in either the desktop or VR condition. Condition assignment was counterbalanced: 10 participants completed the desktop condition, and the other 10 completed the VR condition. After data exclusions, each condition contained eight participants.

\subsection{RMET Task and Experimental Systems}

\subsubsection{RMET Task}

The task was adapted from the Reading the Mind in the Eyes Test (RMET) \cite{baron2001reading}. It consisted of 36 photographs of the human eye region, each accompanied by four descriptive terms. For each item, participants selected the term they believed best represented the mental or emotional state conveyed by the image. One response was recorded for each item. The same 36 items were presented in a fixed order in both conditions. Participants completed the items at their own pace, with no per-question time limit.

\subsubsection{VR and Desktop Implementations}

Two versions of the RMET task were developed in Unity (version 2019.4.7f1): an immersive VR implementation and a desktop implementation. The two implementations presented the same RMET items, response options, layout, item order, and task progression, while differing in presentation platform and interaction hardware.

The VR application was developed using VRTK (version 2.2.1) and presented through an HTC Vive Pro Eye headset. RMET stimuli and response options were displayed on a virtual television within a simulated room. Integrated eye tracking was accessed through the Vive SRanipal API and recorded at 120~Hz. Participants selected their responses using a VR controller. 
In the desktop condition, the RMET was presented on a conventional monitor, and gaze was recorded using a Tobii Eye Tracker 5 at 133~Hz. Participants selected their responses using a mouse.
The VR environment provided an immersive presentation context rather than an interactive social simulation; the RMET stimuli themselves remained two-dimensional images presented on the virtual display. The experimental setups for the VR and desktop conditions are shown in Figure~\ref{fig:Study_Sessions}.


\subsubsection{Multimodal Synchronization}

Brain activity was recorded in both conditions using a Muse 2 EEG headband. Gaze, EEG, participant responses, and task-event timestamps were synchronized using the Lab Streaming Layer (LSL) framework. The \texttt{liblsl-csharp} library was integrated with the Unity applications to stream interaction and gaze events, while EEG was recorded through the corresponding LSL stream. Shared timestamps enabled the multimodal streams to be aligned and segmented into individual RMET question epochs for subsequent feature extraction and predictive modeling.

\subsection{Study Procedure}

At the beginning of the study, participants completed a demographic questionnaire collecting age, gender, educational background, and prior exposure to immersive technologies. Participants were then introduced to their assigned implementation and received instructions on how to navigate the interface and respond to RMET items.
Eye-tracking calibration was performed before the task in both conditions. Participants then completed a practice item to familiarize themselves with the response format and interaction procedure.
Participants completed all 36 RMET items within a single experimental session. The task was organized into two 18-item blocks separated by a participant-controlled rest period with no fixed duration.
After completing the RMET, participants completed the System Usability Scale (SUS) and NASA Task Load Index (NASA-TLX) to report perceived usability and workload, respectively.
The complete experimental session lasted approximately 25-30 minutes, including instructions, calibration, task completion, and the midpoint rest period.

\subsection{Measures and Preprocessing}

\subsubsection{Eye Tracking}

In the desktop condition, gaze was recorded using a Tobii Eye Tracker 5 at 133~Hz and streamed through the Tobii Streaming Engine to LSL. In the VR condition, gaze was recorded using the integrated eye tracker of the HTC Vive Pro Eye at 120~Hz and streamed through a Unity--LSL interface. Both streams were stored in XDF format and subsequently converted to CSV for preprocessing. For both conditions, gaze data were synchronized with the Unity RMET event logs and segmented into question-level epochs using item onset and response timestamps. In VR, gaze rays intersecting the virtual television were projected into normalized two-dimensional screen coordinates. This transformation allowed the same screen-referenced gaze features to be computed in both conditions. However, because the two conditions used different eye-tracking hardware and sampling rates, this common representation does not eliminate device-related measurement differences.
The same fixation--saccade detection algorithm and velocity threshold were applied to both the desktop and VR gaze streams.

Samples with angular velocity below $30^{\circ}$/s were classified as fixation samples, whereas samples above this threshold were classified as saccadic. Continuous sequences of fixation-labeled samples were grouped into fixation periods, with a minimum fixation duration of 100~ms. Samples containing NaN gaze values were discarded before feature extraction. For each question-level epoch, we extracted fixation count, saccade count, and mean fixation time per question. 
To identify mean fixation time per question, we first identified and labeled fixations for each participant in both conditions, then calculated each individual's total fixation time per session. Finally, we computed mean fixation time per question in seconds from the total fixation time at the participant level to account for differences in trial lengths and to provide a participant-level summary of fixation time.

\subsubsection{EEG}

EEG was recorded at 256~Hz using a Muse 2 headband with four dry electrodes positioned at TP9, AF7, AF8, and TP10. Data were acquired through BlueMuse, streamed through LSL, and stored in XDF format before conversion to CSV for preprocessing and alignment with the Unity RMET event logs.

The continuous EEG signal was band-pass filtered from 0.5 to 45~Hz using a zero-phase fourth-order Butterworth filter. The filtered stream was synchronized with the RMET event logs and segmented into question-level epochs using item onset and response timestamps.

EEG preprocessing and artifact rejection were implemented in Python using MNE-Python. Following filtering, epochs were screened using peak-to-peak amplitude thresholding, with epochs exceeding 150~$\mu$V rejected as artifacts.

The auxiliary channel was excluded from feature extraction. For each retained epoch and each of the four EEG channels, absolute band power was computed in five frequency bands, delta (0.5--4~Hz), theta (4--7~Hz), alpha (8--12~Hz), beta (13--30~Hz), and gamma (30--45~Hz), together with spectral entropy.

Absolute band power was estimated from the power spectral density (PSD) using SciPy's Welch procedure and numerical integration over each target frequency range \cite{welch1967use}. The sampling frequency and segment length were both set to 256 samples (\texttt{fs=256}, \texttt{nperseg=256}); the remaining Welch parameters used SciPy's default settings. Spectral entropy was calculated from the normalized PSD using Shannon entropy. No additional offline EEG re-referencing was applied.

\subsubsection{Subjective Measures}

Perceived usability was measured using the System Usability Scale (SUS), a standardized 10-item instrument scored from 0 to 100, with higher scores indicating greater perceived usability \cite{brooke1996sus}.

Perceived workload was assessed using the NASA Task Load Index (NASA-TLX), which evaluates six dimensions of workload: mental demand, physical demand, temporal demand, performance, effort, and frustration \cite{hart1988development}. We used SUS and NASA-TLX outcomes to compare perceived usability and workload between the VR and desktop groups.



\subsection{Predictive Modeling}
\subsubsection{Dataset Construction and Feature Representation}

The predictive dataset was constructed from synchronized EEG, gaze, and interaction streams collected in both the VR and desktop conditions. Data from the two conditions were included in a common modeling dataset. Participant and experimental-condition identifiers were retained as metadata, but the VR/desktop condition label was not included as a predictive feature.

EEG and gaze streams were segmented into question-level epochs using the synchronized Unity event logs. Each epoch was associated with the corresponding participant, experimental condition, RMET item, response, and correctness label. The final question-level dataset contained 576 observations (16 participants $\times$ 36 RMET items). For each question-level epoch, five band-power features and spectral entropy were extracted from each of the four EEG channels, yielding 24 EEG features. The common gaze representation consisted of three features: fixation count, saccade count, and mean fixation time per question.

Although the VR eye tracker produced additional signals, including pupil diameter, eye openness, three-dimensional gaze direction, and headset/controller information, these VR-specific variables were not included in the predictive models. Only gaze features that could be derived consistently in both VR and desktop conditions were retained. The same fixation--saccade detection algorithm and threshold were applied to gaze data from both platforms. The EEG and gaze features were aligned by question and combined to form the multimodal feature representation used for predictive modeling.

\subsubsection{Model Preprocessing}

Before model fitting, input features were standardized to zero mean and unit variance using \textit{}{StandardScaler}. All preprocessing was performed independently within each cross-validation training fold: scaling parameters were estimated from the training data and subsequently applied to the corresponding held-out data.

Principal Component Analysis (PCA) was subsequently applied for dimensionality reduction, retaining components that explained 95\% of the variance. PCA was fitted only on the training portion of each cross-validation iteration and then used to transform the corresponding held-out data. For the EEG-only, gaze-only, and combined EEG+gaze analyses, PCA was fitted separately using only the features included in that specific input configuration.

\subsubsection{Question-Level Classification}

The classification task predicted whether each RMET response was correct ($1$) or incorrect ($0$) from the features extracted during that question. Five classifiers were evaluated: Logistic Regression (LR), Support Vector Machine (SVM), Random Forest (RF), XGBoost, and LightGBM. Classification was evaluated using five-fold StratifiedGroupKFold cross-validation, with participant identity used as the grouping variable and response correctness (correct/incorrect) used for stratification. All question-level observations from a participant therefore remained within the same fold, preventing participant-level leakage between training and test data.

Feature standardization and PCA were fitted exclusively on the training portion of each fold and applied to the corresponding held-out participants. 
Random Forest was configured with 50 trees, and XGBoost used a learning rate of 0.1. Classification performance was evaluated using accuracy, precision, recall, and F1 score. Precision, recall, and F1 were calculated as binary metrics, with correct responses ($1$) treated as the positive class. 

\subsubsection{Participant-Level Score Regression}

A separate regression analysis was conducted to predict each participant's total RMET score. For each participant, the 36 question-level feature vectors were concatenated using a flattening operation to form a single one-dimensional participant-level representation. Thus, question-level features were retained separately rather than averaged across questions. The participant's total RMET score was calculated by summing correct responses across the 36 items and served as the regression target. For the combined EEG+gaze configuration, each question contained 27 features (24 EEG and three gaze), yielding 36 $\times$ 27 = 972 input features per participant before scaling and PCA. We retained question-specific features rather than averaging across items so that the representation preserved variation associated with individual RMET stimuli. This choice substantially increases dimensionality.

Four ensemble regression models were evaluated: Random Forest, XGBoost, Gradient Boosting, and Bagging Regressor. All four regression models were configured with 100 estimators. XGBoost and Gradient Boosting additionally used a learning rate of 0.1 and a maximum tree depth of 4. No data-driven hyperparameter tuning was performed.

Regression performance was evaluated using leave-one-participant-out cross-validation (LOOCV). In each iteration, the model was trained on 15 participants and evaluated on the remaining participant, maintaining complete participant-level separation between training and test data.

Within each LOOCV iteration, feature standardization and PCA were fitted using only the training participants and then applied to the held-out participant. Regression performance was evaluated using mean absolute error (MAE), root mean squared error (RMSE), and $R^2$.

\subsubsection{Modality Ablation and Model Interpretation}

To examine the predictive contribution of each sensing modality, the regression pipeline was evaluated under three input configurations: EEG-only, gaze-only, and combined EEG+gaze. Scaling and PCA were re-fitted independently for each modality configuration within each LOOCV iteration.

\begin{figure*}[t]
  \centering
  \includegraphics[width=\textwidth]{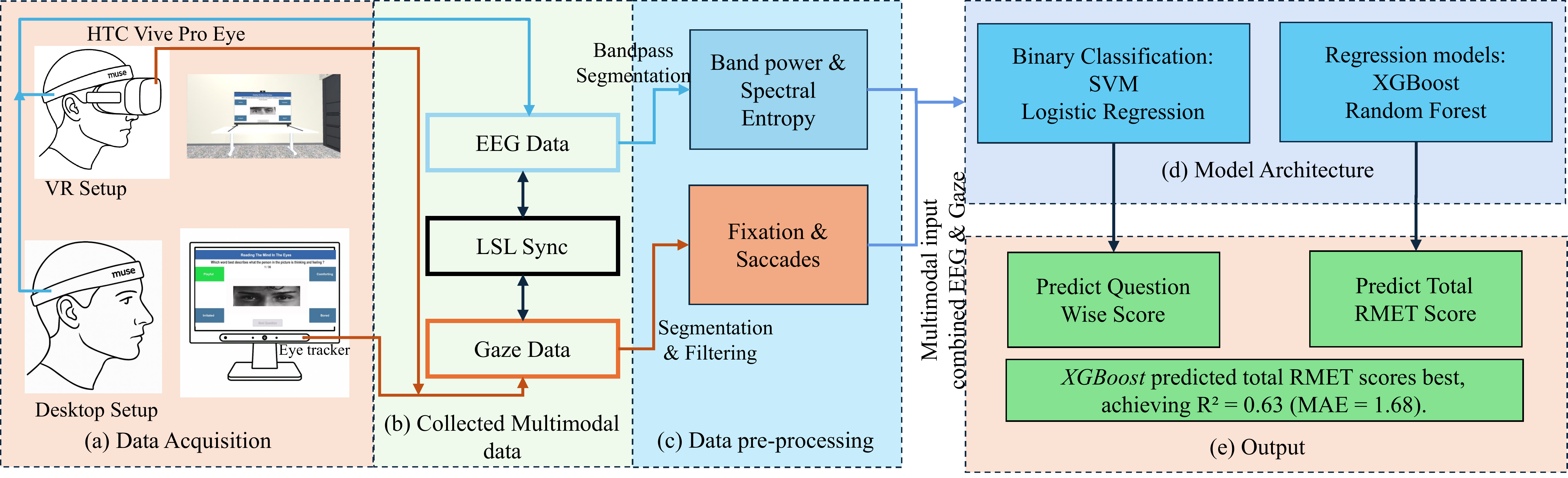}
  \caption{Overview of the multimodal data pipeline for score prediction using EEG and gaze streams. (a) Data Acquisition: EEG signals are recorded using a Muse headband, while gaze data is collected via the HTC Vive Pro Eye Tracker (VR) or the Tobii Eye Tracker (Desktop) during RMET tasks. (b) Collected Data: Raw EEG and gaze data are collected and synchronized using the Lab Streaming Layer. (c) Data Pre-processing: EEG data undergo bandpass segmentation followed by the extraction of bandpower and spectral entropy features. Gaze data are filtered and analyzed to extract fixation and saccade counts. (d) Model Architecture: Multiple machine learning models are trained on the preprocessed features using cross-validation methods. (e) Output: The trained models predict question correctness or the total RMET scores of a participant.}
  \label{fig:teaser}
\end{figure*}

The overall pipeline for EEG- and gaze-based RMET performance prediction is illustrated in Figure~\ref{fig:teaser}.

\section{Results and Findings} \label{sec:results}
\subsection{Machine Learning Model Evaluation}

To classify whether each question was answered correctly, we evaluated our dataset using a range of machine-learning models. The binarized classification task reflects a class distribution of approximately 65\% correct and 35\% incorrect responses. This mild imbalance may inflate accuracy and recall, especially in models biased toward the majority class (i.e., predicting more correct responses). To mitigate this, we reported multiple metrics, including precision, recall, and F1-score, which better capture model performance under imbalance. Table~\ref{tab:model_comparison} summarizes the performance based on 5-fold cross-validation. \textbf{SVM} achieved perfect recall (1.00) and the highest F1-score (0.81), indicating its strong ability to identify positive cases, although its precision was moderate. \textbf{XGBoost} and \textbf{Random Forest} had the same F1-scores of 0.77, and demonstrated high recall values above 0.80. \textbf{LightGBM} also performed competitively with an F1-score of 0.75, though its accuracy was slightly lower. \textbf{Logistic Regression}, on the other hand, yielded the lowest accuracy (0.53) and F1-score (0.59), with noticeable variation across folds. These results highlight the relative effectiveness of ensemble-based models and SVM in handling the classification task for this dataset, while simpler models like Logistic Regression struggled with both precision and consistency.

We further explored the prediction of participants’ total scores using regression models and assessed the contribution of each modality through an ablation study. As shown in Table \ref{tab:ablation_study}, across all four evaluated regressors, the combined EEG+gaze representation produced lower observed Mean Absolute Error (MAE), Root Mean Squared Error (RMSE), and coefficient of determination ($R^2$). Among all models, \textbf{XGBoost} achieved the strongest observed metrics in the combined setting, with the lowest MAE (1.68) and RMSE (1.96), and the highest $R^2$ (0.63). 
\textbf{Gradient Boosting} also performed relatively well in the combined setting (MAE = 2.13, RMSE = 2.64, $R^2$ = 0.32), whereas \textbf{Random Forest} (MAE = 2.25, RMSE = 2.69, $R^2$ = 0.29) and \textbf{Bagging} (MAE = 2.33, RMSE = 2.83, $R^2$ = 0.22) showed comparatively weaker performance. All EEG-only and gaze-only configurations produced negative held-out \(R^2\) values, indicating that these unimodal predictions performed worse than a mean-reference prediction according to this metric. For example, XGBoost’s \(R^2\) improved from \(-1.04\) with EEG only and \(-0.70\) with Gaze only to \(0.63\) with combined EEG+Gaze features, with corresponding reductions in both MAE and RMSE. This consistent descriptive pattern motivates further investigation of whether the two modalities provide complementary predictive information.

\begin{table}[ht]
    \centering
    \caption{Performance comparison of machine learning models in question correctness prediction as a binary classification problem. Higher values are better. \textbf{Machine Learning Models:} Logistic Regression (LR), Support Vector Machine (SVM), Random Forest (RF), XGBoost, and LightGBM. \textbf{Metrics:} Accuracy, Precision, Recall, and F1 Score }
    \label{tab:model_comparison}
    \footnotesize
    \begin{tabular}{lcccc}
        \hline
        \textbf{Model} & \textbf{Accuracy} & \textbf{Precision} & \textbf{Recall} & \textbf{F1 Score} \\
        \hline
        LR        & 0.53 & \purple{0.70} & 0.52 & 0.59 \\
        SVM       & \textbf{\purple{0.68}} & \textbf{0.67} & \textbf{1.00} & \textbf{\purple{0.81}} \\
        RF        & \purple{0.67} & \textbf{\purple{0.72}} & \purple{0.84} & \purple{0.77} \\
        XGBoost   & \purple{0.66} & 0.69 & \purple{0.87} & 0.77 \\
        LightGBM  & \purple{0.64} & \purple{0.70} & 0.81 & 0.75 \\
        \hline
    \end{tabular}
\end{table}

\begin{table*}[ht]
    \centering
    \caption{Ablation Study for the Regression Models. \textbf{Machine Learning Models:} Random Forest (RF), XGBoost, Gradient Boosting, and Bagging. \textbf{Metrics:} Mean Absolute Error (MAE), Root Mean Squared Error (RMSE), and Coefficient of Determination ($R^2$). Lower MAE and RMSE, and higher $R^2$ values indicate better performance. \textbf{Features:} EEG Only, Gaze Only, and Combined EEG + Gaze.}
    \label{tab:ablation_study}
    \footnotesize
    \begin{tabular}{lccccccccc}
        \hline
        \multirow{2}{*}{\textbf{Model}} &
        \multicolumn{3}{c}{\textbf{EEG Only}} &
        \multicolumn{3}{c}{\textbf{Gaze Only}} &
        \multicolumn{3}{c}{\textbf{Combined}} \\
        \cline{2-10}
         & \textbf{MAE} & \textbf{RMSE} & \textbf{$R^2$} &
           \textbf{MAE} & \textbf{RMSE} & \textbf{$R^2$} &
           \textbf{MAE} & \textbf{RMSE} & \textbf{$R^2$} \\
        \hline
        Random Forest
            & \purple{2.86} & \purple{3.49} & \purple{-0.19}
            & \purple{3.39} & \purple{3.89} & \purple{-0.48}
            & \purple{2.25} & \purple{2.69} & 0.29 \\
        XGBoost
            & 3.87 & 4.57 & -1.04
            & 3.61 & 4.17 & -0.70
            & \textbf{1.68} & \textbf{1.96} & \textbf{0.63} \\
        Gradient Boosting
            & \purple{4.66} & \purple{5.13} & \purple{-1.57}
            & \purple{3.98} & \purple{4.59} & \purple{-1.05}
            & \purple{2.13} & \purple{2.64} & \purple{0.32} \\
        Bagging
            & \textbf{\purple{2.79}} & \textbf{\purple{3.45}} & \textbf{\purple{-0.16}}
            & \textbf{\purple{3.31}} & \textbf{\purple{3.74}} & \textbf{\purple{-0.36}}
            & \purple{2.33} & \purple{2.83} & \purple{0.22} \\
        \hline
    \end{tabular}
\end{table*}



\subsection{Eye-tracking Analysis}
To characterize gaze behavior across the VR and desktop groups, we compared participant-level mean fixation time per question. 
The VR condition resulted in a slightly higher mean fixation time per question (in seconds) ($M=4.16$, $SE=0.94$) than the desktop condition ($M=3.82$, $SE=0.46$), but this difference was not statistically significant, $t(14)=0.33$, $p=.749$. 
\subsection{Subjective Measures}
 
 We performed independent-samples $t$-tests to compare the VR and desktop groups on SUS scores, overall Raw NASA-TLX scores, and the six NASA-TLX dimensions. Dimension-level NASA-TLX comparisons were exploratory and were not adjusted for multiple comparisons. Results are reported as mean ($M$) and standard error ($SE$).
 
\begin{figure}[ht]
    \centering
    \includegraphics[width=1\linewidth]{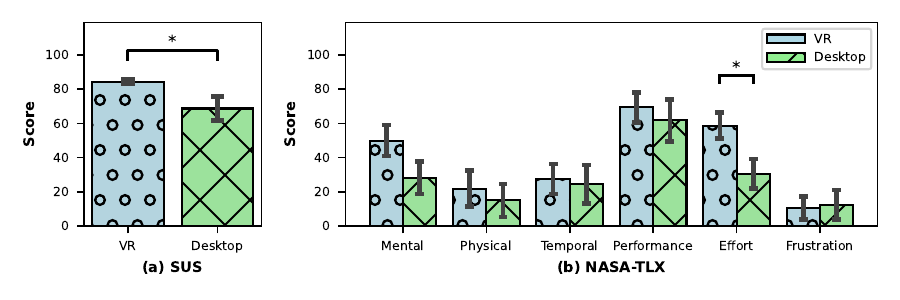}
   \caption{Questionnaire results: (a) SUS score: higher SUS scores indicate better usability and (b) NASA-TLX: lower scores show a reduced task load in accomplishing the task. Combined SUS scores showed significant differences between VR group and Desktop group users, favoring VR. Also, in the effort category, the VR group presented higher values than the Desktop group.}
       \label{fig:sus_nasa}
\end{figure}

\textbf{Usability (SUS):}
The VR group reported higher perceived usability ($M=84.38$, $SE=1.40$) than the desktop group ($M=68.75$, $SE=7.13$), $t(14)=2.15$, $p=.049$, as shown in ~\autoref{fig:sus_nasa}(a).
 
\textbf{Task Load (NASA-TLX):} 
Overall Raw NASA-TLX scores were descriptively higher in the VR group ($M=39.69$, $SE=3.51$) than in the desktop group ($M=28.75$, $SE=7.34$), but the difference was not statistically significant, $t(14)=1.34$, $p=.200$. Among the six individual NASA-TLX dimensions, Mental, Physical, Temporal, Performance, and Frustration did not differ significantly between groups (all $p>.05$), as shown in ~\autoref{fig:sus_nasa}(b). Effort was higher in the VR group ($M=58.75$, $SE=7.72$) than in the desktop group ($M=30.62$, $SE=8.53$), $t(14)=2.45$, $p=.028$. This dimension-level comparison was exploratory and unadjusted for multiple comparisons.

\section{Discussion}
\label{sec:discussion}

This study examined multimodal prediction of RMET performance alongside the experience of completing the assessment in VR or on a desktop. Combined EEG--gaze inputs yielded better observed total-score prediction than either modality alone across all four regressors. The platform comparison revealed higher perceived usability and effort in VR, with no statistically significant difference in mean fixation time per question. These findings inform both the development of performance models from synchronized sensing and the design of interfaces through which assessment data are collected.

\subsection{RQ1: Multimodal Prediction of RMET Performance}

The clearest predictive pattern emerged from the participant-level regression analysis. Across Random Forest, XGBoost, Gradient Boosting, and Bagging, combining EEG and gaze reduced MAE and RMSE and increased $R^2$ relative to either modality alone. XGBoost achieved the strongest observed performance with combined inputs ($R^2=0.63$, MAE $=1.68$, RMSE $=1.96$). All unimodal configurations produced low $R^2$ values, whereas all combined configurations produced higher values. The consistency of this pattern across regressors supports further investigation of the combined representation, although its magnitude and stability require confirmation in larger samples.

This pattern is consistent with previous evidence that EEG and eye-tracking measurements can provide complementary predictive information. In emotion recognition, combining EEG with pupillary responses improved classification relative to either modality alone~\cite{Zheng2014}. However, benefits are sensitive to how the modalities are represented and combined. In a comparison of EEG and eye-tracking approaches for attention classification, late fusion outperformed both unimodal approaches, whereas early fusion did not consistently improve performance~\cite{vortmann2022multimodal}. Our study examines this possibility for RMET score prediction using EEG spectral features and gaze measures. Differences in targets, sensors, features, and validation procedures preclude direct numerical comparisons with those studies. The observed improvement is compatible with non-redundant information in the two modalities, but does not identify the cognitive processes responsible for that improvement.

The question-level classification results should be interpreted cautiously. SVM achieved the highest reported F1 score and a recall of 1.00, indicating high sensitivity to correct responses. However, approximately 65\% of responses were correct, and strong performance on this class does not necessarily indicate reliable identification of incorrect responses. Given the class distribution and small participant sample, these findings remain exploratory.

These analyses used pooled VR and desktop data, with participants separated between training and testing. They consequently evaluate prediction for held-out participants within the combined study sample. They do not establish whether VR improves predictive performance or whether a model trained on one platform transfers to the other. Larger datasets would permit separate assessment of modality benefits within each platform and explicit tests of transfer between platforms.

\subsection{RQ2: VR and Desktop Assessment Experience}

Participants in the VR condition reported higher SUS usability scores, whereas overall Raw NASA-TLX workload did not differ significantly between groups. Among the six exploratory, unadjusted NASA-TLX dimension comparisons, the VR group reported higher effort. The other five NASA-TLX dimensions did not differ significantly between conditions. The observed pattern illustrates that favorable usability ratings need not coincide with lower reported effort. The effort result concerns one workload dimension, and the small groups and multiple workload comparisons warrant an exploratory interpretation.

Related VR research illustrates why experience, task demands, and performance should be evaluated separately. In a science-learning simulation, immersive VR increased presence while increasing EEG-derived cognitive load and reducing learning relative to desktop presentation~\cite{makransky2019adding}. Our assessment differs in both task and measures, so that result cannot explain the higher effort reported here. It nevertheless provides a useful comparison that favorable experience ratings and performance-related outcomes need not change together. For assessment design, higher usability should therefore be considered alongside the demands introduced by the interface.

Mean fixation time per question did not differ significantly between platforms. This finding does not establish equivalent gaze behavior, particularly with eight participants per condition, and a single aggregate gaze measure cannot characterize engagement comprehensively. Similarly, SUS measures perceived usability rather than engagement or presence. The results support a difference in reported experience without establishing that VR increased attention, reduced distraction, or improved RMET performance.

\subsection{Implications for AI-Supported VR Assessment}

The contribution at the intersection of AI and VR is the integration of a controlled assessment, synchronized behavioral and physiological recordings, and participant-wise performance modeling. RMET responses already permit direct calculation of correctness and total scores. Because RMET correctness and total scores can already be computed directly from responses, the present models are not intended to replace conventional scoring. Instead, they test whether synchronized behavioral and physiological signals contain information associated with task performance, a prerequisite for future work on user models that operate during interaction rather than only after task completion. Establishing practical benefit would require showing that these models support useful decisions during an assessment or training session.

The VR implementation retained the two-dimensional RMET images on a virtual display. This preserved common assessment content across platforms and supported comparison of their delivery, while limiting the social interaction represented by the task. The VR implementation should therefore be interpreted as immersive delivery of a conventional visual assessment rather than as a socially interactive VR task. Prior validation of VR-EAL combined comparisons with established cognitive measures and evaluation of the immersive assessment experience~\cite{kourtesis2021validation}. A similar distinction between predictive performance, assessment validity, and user experience is needed here.

The prediction target also constrains interpretation. Concerns about the RMET's relationship to emotion recognition and broader theory-of-mind abilities support treating its scores as task-specific outcomes~\cite{oakley2016theory}. EEG and gaze may contain information associated with those outcomes without constituting a general measure of social intelligence. Additional tasks and independent validation would be needed to establish whether the observed predictive patterns extend to other aspects of social cognition.

\subsection{Limitations}

This study has several limitations. The small and relatively homogeneous sample (\(N=16\)) limits the stability and generalizability of the predictive results, despite participant-wise cross-validation. The VR and desktop groups used different eye-tracking devices, so between-group differences cannot be attributed to the presentation platform alone. The EEG preprocessing may also have left some movement or other artifacts in the signals. In addition, the regression models used many features relative to the number of participants and should therefore be considered exploratory. Finally, the VR task presented static RMET images on a virtual display rather than a dynamic social interaction, limiting conclusions beyond this specific task.
\subsection{Future Directions}

Future studies should evaluate the approach with larger, more diverse samples, independent validation data, and randomized platform comparisons. Testing additional social-cognitive tasks and recording systems would help establish whether the predictive patterns generalize. Interactive VR scenarios involving virtual agents, dynamic social cues, and speech could extend the approach beyond static images. After establishing predictive validity, studies could examine inference during task completion and whether model-informed adaptations improve assessment usability and usefulness while preserving measurement validity.

\section{Conclusion}\label{sec:conclusion}
This proof-of-concept study examined AI-based modeling of task-specific RMET performance using synchronized EEG and gaze collected in VR and desktop settings. Across the evaluated regressors, combined EEG--gaze representations showed lower observed prediction errors than either modality alone, although the small sample limits the stability and generalizability of these estimates. The VR group reported higher usability, while overall workload and mean fixation time per question did not differ significantly; an exploratory unadjusted comparison indicated higher reported effort in VR. These findings motivate larger studies to validate the observed multimodal prediction pattern and further examine assessment experience across immersive and conventional settings.

\section*{Acknowledgment} 
The authors used ChatGPT to improve the clarity and structure of the manuscript. All content was reviewed and edited by the authors, who take full responsibility for the final manuscript.


\bibliographystyle{IEEEtran}
\bibliography{Main_Paper}

\end{document}